\documentclass[12pt,letterpaper]{article} 
\pdfoutput=1

\usepackage[includeheadfoot,
            marginratio={1:1,2:3}, 
            width=412pt, 
            height=688pt,]{geometry}

\usepackage{amsmath}
\usepackage{amsfonts}
\usepackage{amssymb}
\usepackage{stmaryrd}
\usepackage{graphicx}
\usepackage{cite}

\newcommand{\nc}{\newcommand}
\nc{\lb}{\llbracket}
\nc{\rb}{\rrbracket}
\nc{\gl}{\llbracket}
\nc{\gr}{\rrbracket}

\newcommand{\eq}[1]{\begin{equation}
                     \begin{split} #1 \end{split}
                     \end{equation}}

\newcommand{\ov}{\overline}

\newcommand{\drawsquare}[2]{\hbox{% 
\rule{#2pt}{#1pt}\hskip-#2pt%  left vertical 
\rule{#1pt}{#2pt}\hskip-#1pt%  lower horizontal 
\rule[#1pt]{#1pt}{#2pt}}\rule[#1pt]{#2pt}{#2pt}\hskip-#2pt%  upper horizontal 
\rule{#2pt}{#1pt}}% right vertical 
\newcommand{\fund}{\raisebox{-.5pt}{\drawsquare{6.5}{0.4}}}%  fund 
\newcommand{\Ysymm}{\raisebox{-.5pt}{\drawsquare{6.5}{0.4}}\hskip-0.4pt% 
         \raisebox{-.5pt}{\drawsquare{6.5}{0.4}}}%  symmetric second rank 
\newcommand{\Yasymm}{\raisebox{-3.5pt}{\drawsquare{6.5}{0.4}}\hskip-6.9pt% 
        \raisebox{3pt}{\drawsquare{6.5}{0.4}}}%  antisymmetric second rank 
\newcommand{\antifund}{\overline{\fund}}

\def\i{\iota} 

\allowdisplaybreaks[2]
\numberwithin{equation}{section}

\begin{document}

\vspace*{-1.5cm}
\begin{flushright}
  {\small
  MPP-2015-292\\
  }
\end{flushright}

\vspace{1.5cm}
\begin{center}
{\LARGE Intersecting Branes, SUSY Breaking \\[0.2cm]
              and   the  $2\,$TeV  Excess at the LHC}
\vspace{0.4cm}

\end{center}

\vspace{0.35cm}
\begin{center}
  Ralph Blumenhagen
\end{center}

\vspace{0.1cm}
\begin{center} 
\emph{Max-Planck-Institut f\"ur Physik (Werner-Heisenberg-Institut), \\ 
   F\"ohringer Ring 6,  80805 M\"unchen, Germany } \\[0.1cm] 
\vspace{0.25cm}

\vspace{0.2cm}

 \vspace{0.5cm} 
\end{center} 

\vspace{1cm}

%%%%%%%%%%%%%%%%%%%%%%%%%%%%%%%%%%%%%%%%%%%%%%%
%%%%%%%%%%%%%%%%%%%%%%%%%%%%%%%%%%%%%%%%%%%%%%%
%%%%%%%%%%%%%%%%%%%%%%%%%%%%%%%%%%%%%%%%%%%%%%%
%%%%%%%%%%%%%%%%%%%%%%%%%%%%%%%%%%%%%%%%%%%%%%%
%%%%%%%%%%%%%%%%%%%%%%%%%%%%%%%%%%%%%%%%%%%%%%%
%%%%%%%%%%%%%%%%%%%%%%%%%%%%%%%%%%%%%%%%%%%%%%%
%%%%%%%%%%%%%%%%%%%%%%%%%%%%%%%%%%%%%%%%%%%%%%%
%%%%%%%%%%%%%%%%%%%%%%%%%%%%%%%%%%%%%%%%%%%%%%%

\begin{abstract}
Intersecting D-brane models in string theory can naturally support the gauge
and matter content of left-right symmetric extensions
of the Standard Model with gauge symmetry $SU(3)_c\times
SU(2)_L\times SU(2)_R\times U(1)_{B-L}$.  Considering such models
as   candidates for explaining the $2\,$TeV excesses seen 
in Run-1 by both ATLAS and CMS, the minimal possible scale of supersymmetry breaking
is determined by  the requirement of precise one-loop gauge coupling unification.
For the  vector-like, bifundamental and (anti-)symmetric Higgs content
of  such brane configurations, 
this comes out  fairly universally
at around $19\,$TeV.  For the $SU(2)_R$ gauge
coupling one finds values $0.48<g_R(M_R)< 0.6$. Threshold corrections
can potentially lower the scale of supersymmetry breaking.
\end{abstract}

\clearpage

%\tableofcontents

%%%%%%%%%%%%%%%%%%%%%%%%%%%%%%%%%%%%%%%%%%%%%%%
%%%%%%%%%%%%%%%%%%%%%%%%%%%%%%%%%%%%%%%%%%%%%%%
%%%%%%%%%%%%%%%%%%%%%%%%%%%%%%%%%%%%%%%%%%%%%%%
%%%%%%%%%%%%%%%%%%%%%%%%%%%%%%%%%%%%%%%%%%%%%%%
%%%%%%%%%%%%%%%%%%%%%%%%%%%%%%%%%%%%%%%%%%%%%%%
%%%%%%%%%%%%%%%%%%%%%%%%%%%%%%%%%%%%%%%%%%%%%%%
%%%%%%%%%%%%%%%%%%%%%%%%%%%%%%%%%%%%%%%%%%%%%%%
%%%%%%%%%%%%%%%%%%%%%%%%%%%%%%%%%%%%%%%%%%%%%%%

\section{Introduction}

In the 8 TeV run of the LHC the Higgs particle was discovered, which
completes the particle spectrum of the Standard Model.
Concerning physics beyond the SM, the hope was that
supersymmetry would be found at a scale not far above the weak
scale, thus providing a solution to the hierarchy problem.
Even though no  direct indication of supersymmetry has appeared yet,   
some anomalies were reported indicating with  2-3$\sigma$  a resonance
of around $2\,$TeV in the di-boson decay channel by ATLAS \cite{Atlasexcess}
and in the  $e^+\, e^-\, j\, j$, $W\, h^0$ and $j\,j$ final states  by  CMS \cite{CMSexcess}. 
The most significant one is the ATLAS 3.4$\sigma$ excess in the 
hadronic decay of a $WZ$ pair of electro-weak gauge bosons.
 
Promising candidates
to explain these excesses are left-right symmetric extensions of
the Standard Model (LRSM) \cite{Mohapatra:1974hk}  with gauge symmetry
$SU(3)_c\times SU(2)_L\times SU(2)_R\times U(1)_{B-L}$
in which a bi-doublet Higgs field
naturally generates mixings between the left and right $SU(2)$ 
W-bosons. The breaking of $SU(2)_R\times U(1)_{B-L}$ to the Standard Model
hypercharge   at the scale
$M_R\sim 2\,$TeV  can be performed e.g. by
an $SU(2)_R$ triplet   or simply by a doublet.
The phenomenological potential of such models
to explain all the excesses seen by ATLAS and CMS were e.g.
analyzed in \cite{phenoLR}. In these papers, the $SU(2)_R$  gauge coupling
was restricted to be in the regime $0.4<g_R(M_R)<0.6$. 

Taking the precision measurements for dilepton channels into account,
some authors came to the conclusion that  one should better use 
a leptophobic version of the LRSM, where the right handed
leptons are not sitting in a doublet of $SU(2)_R$. 
Formally, this makes the model less natural so that
our attitude is that both  experimentally and phenomenologically this 
 issue is not yet completely settled so that here  we proceed  to 
consider  the standard version of an LRSM. 
Then, right-handed neutrinos and the  $Z'$ gauge boson should have  masses
larger than $M_{W'}$.

In fact, in this letter we are not so much concerned with such  phenomenological
fine-print, but would like to evaluate the
potential of string theory inspired LRSM-like models to provide 
the overarching structure in the regime between $M_R$ and
a unification scale $M_U$. Let us mention that a string inspired explanation
for the $2\,$TeV excesses was already proposed in terms of 
an anomalous $Z'$ gauge boson in \cite{Anchordoqui:2015uea}
(see also \cite{Feng:2015second} for application to LHCb $b\to s\ell^+\ell^-$ anomalies).
Note that there the mass of the anomalous $Z'$ is of the order of the  
string scale $M_S$, which then is bound to be small.

In this letter we follow a different route and consider stringy realizations of LRSMs.
One option is to embed the LRSM into a grand unified gauge group $SO(10)$, 
as it can be realized in heterotic or F-theory compactifications.
In these constructions, one often gets
a bunch of extra particles that can  generate fast proton decay
and  one needs to solve the problem of mass hierarchies inside  the
Higgs multiplets (like the doublet-triplet splitting problem for SU(5)
GUTs).

Since all the fields in the LRSM sit  in bifundamental 
and (anti-)symmetric 
representations of the gauge group, it is also very natural
to realize such models from intersecting D-branes \cite{IBWorlds}
(see \cite{Blumenhagen:2006ci} for reviews), thereby avoiding
some of the above mentioned  problems present in GUT models. 
Baryon number is known to be  a perturbative global symmetry in these
constructions protecting the proton from a too fast decay.
Therefore, in this paper we consider LRSM realizations
by intersecting branes focusing on  D7-branes in type IIB orientifolds.   

Since in controllable models of string
theory,  supersymmetry needs to be broken at a scale $M_{\rm SUSY}$
smaller than the string scale $M_S\sim M_U$, we consider the hierarchy of
mass scales $M_Z<M_R<M_{\rm SUSY}<M_U$ and analyze the issue
of gauge coupling unification.  Such an analysis has been performed
in a field theory context (see e.g. \cite{LRSMunify,Hirsch:2015fvq} for more recent  studies), but to our knowledge it was always assumed
either that one does not have supersymmetry below the GUT scale at all or that the
intermediate hierarchy of scales is reversed, i.e.  $M_Z<M_{\rm
  SUSY}<M_R<M_U$.
Thus, the observations made in this letter can be considered as complementary to
earlier results reported in the literature.

This letter is organized as follows:
In section 2, we review a few important aspects of  intersecting
D-brane models where in particular we discuss under what 
circumstance one can still get gauge coupling unification at the
string scale. Section 3
provides two simple prototype examples
of possible realizations of a LRSM. 

In section 4 the one-loop
running of the four gauge couplings is analyzed. 
Note that this analysis only depends on the  matter content of the models
and is therefore  generically valid, i.e. without necessarily referring
to an intersecting D-brane scenario.
After fixing the Higgs sector, requiring
precise one-loop gauge coupling unification and using $M_R=2\,$TeV, 
one can uniquely determine the unification scale,
the supersymmetry breaking scale and the $SU(2)_R$ gauge
coupling $g_R$. It is observed that the minimal possible value
for the supersymmetry breaking scale, $19\,$TeV, shows a certain universal
behavior. Moreover, the value of the $SU(2)_R$ gauge coupling
come out as $0.48<g_R(M_R)< 0.6$, depending on the 
vector-like Higgs sector.
\vspace{0.1cm}

%Even though  a more thorough investigation of the phenomenological
%details is beyond the scope of this letter, 
%we consider these observations as complementary to
%earlier results reported in the literature.
{\it Note added}: The latest announcement of physics results by the
ATLAS+CMS collaborations on 15.12.2015 did not show any $2\,$TeV excesses
at Run-2. This of course diminishes the experimental motivation for
the analysis done in this letter, but we think that the general
results on the relation of the scales in LRSMs and supersymmetry that
are  obtained are nevertheless interesting.

%%%%%%%%%%%%%%%%%%%%%%%%%%%%%%%%%%%%%%%%%%%%%%%
%%%%%%%%%%%%%%%%%%%%%%%%%%%%%%%%%%%%%%%%%%%%%%%
%%%%%%%%%%%%%%%%%%%%%%%%%%%%%%%%%%%%%%%%%%%%%%%%%%%%%%%%%%%%%%%%%%%%%%%%%%%%%%%%%%%%%%%%%%%%%%%%%%%
%%%%%%%%%%%%%%%%%%%%%%%%%%%%%%%%%%%%%%%%%%
%%%%%%%%%%%%%%%%%%%%%%%%%%%%%%%%%%%%%%%%%%%%%%%
%%%%%%%%%%%%%%%%%%%%%%%%%%%%%%%%%%%%%%%%%%%%%%%
%%%%%%%%%%%%%%%%%%%%%%%%%%%%%%%%%%%%%%%%%%%%%%%

\section{Models of intersecting D7-branes}

Let us review some of the main ingredients for the construction of
intersecting D7-brane models in orientifolds of the Type IIB
superstring. For more details we refer to the
reviews \cite{Blumenhagen:2006ci} and in particular to \cite{Blumenhagen:2008zz}.

One considers the Type IIB superstring compactified on a 
Calabi-Yau threefold $X$ and performs the quotient 
by an orientifold projection $\Omega\sigma (-1)^{F_L}$, where
$\Omega$ is the world-sheet parity reversal and $\sigma$ 
denotes a holomorphic involution of the Calabi-Yau 
satisfying $\sigma(J)=J$ and $\sigma(\Omega_3)=-\Omega_3$.
Here $J$ is the K\"ahler form and $\Omega_3$ is the holomorphic
$(3,0)$ form on the threefold. At the fixed locus of $\sigma$,
one gets O7 and O3-planes, whose R-R charges (tadpoles)
need to be cancelled by the introduction of D7 and D3-branes.

Stacks of $N_a$  D7-branes can wrap holomorphic  four-cycles $\Sigma_a$ in
the homology class $H_4(X,\mathbb Z)$ of the threefold.
Moreover, these D7-branes can carry a non-trivial gauge flux 
determining a line bundle ${\cal L}_a$ on $\Sigma_a$.
Depending on whether this D7-brane configuration is invariant
under the orientifold projection or not, one gets $SO(N)/SP(N)$-gauge groups
or $U(N)$ gauge groups, respectively. Note that one gets
$U(N)=SU(N)\times U(1)$ instead of $SU(N)$. 

The most simple and phenomenologically useful  configurations for
SM-like model building arise for branes wrapping
orientifold invariant,
rigid four-cycles $\Sigma_a$ of $SP$-type with the different stacks being
only distinguished by their gauge flux ${\cal L}_a$.
If $c_1({\cal L}_a)\in H^2_-(\Sigma_a,\mathbb Z)$ then such a  stack carries
gauge group $SP(N_a)$ and for a flux $c_1({\cal L}_a)\in H^2_+(\Sigma_a,\mathbb Z)$
one gets an orientifold image line bundle  ${\cal L}'_a$ so that
together they support a  $U(N_a)$ gauge group.
 
The massless  modes between two such stacks of branes transform in  
bifundamental or (anti)-symmetric  representations, where the chiral index is given by
\eq{
                      I_{ab}=  -\int_{X} [\Sigma_a]\wedge [\Sigma_b]\wedge
                      \big( c_1(L_a)-c_1(L_b)\big)\,.
}      
Here $[\Sigma]$ denotes the Poincar\'e dual two form to the four-cycle
$\Sigma$. For $U(N)$ stacks of branes the various open string
sectors give the chiral spectrum summarized in table \ref{tab_chir_spec}.
\begin{table}[htbp] 
  \renewcommand{\arraystretch}{1.5} 
  \begin{center} 
    \begin{tabular}{|c|c|c||c|} 
      \hline 
      \hline 
      sector & $U(N_a)$ & $U(N_b)$ & chirality   \\ 
      \hline \hline 
      $(a b)$         & $\antifund_{\, (-1)}$ & $\fund_{\, (1)}$  & $I_{ab}  $ \\
      \hline
      $(a' b)$         & $\fund_{\, (1)}$ & $\fund_{\, (1)}$   &  $ I_{a'b}$  \\
      \hline 
      $(a'a)$         & $\Yasymm_{\,(2)}$ & $1$ & $\frac{1}{2}( I_{a'a}+2I_{\rm{O7}a})   $  \\ 
      \hline
      $(a'a)$         & $\Ysymm_{\,(2)} $ & $1$ & $\frac{1}{2}( I_{a'a}-2I_{\rm{O7}a})   $ \\
      \hline 
      \hline  
    \end{tabular} 
    \caption{Chiral spectrum for intersecting D7-branes. The
      subscripts denote $U(1)$ charges.}
    \label{tab_chir_spec} 
  \end{center} 
\end{table}

\vspace{-0.5cm}
\noindent
For an $SP$-stack, chiral fields can only arise from the intersection
with a $U(N)$ stack.

The non-chiral (vector-like) part of the spectrum can be determined
by computing certain line bundle cohomology groups on $\Sigma$
\footnote{For concrete model building  it is important to take into account that the
four-cycle $\Sigma$ can contain homological  two-cycles that are not 
closed in the ambient threefold $X$. One can always twist
a line bundle ${\cal L}_a$ by such a trivial bundle ${\cal R}_a$ without 
changing the chiral spectrum. However, the vector-like part
computed via \eqref{externcoh} for the bundle  ${\cal L}_a\otimes {\cal R}_a$ will change.}.
For branes wrapping the same four-cycle $\Sigma$, this was
determined in \cite{Blumenhagen:2008zz} and for completeness we
provide the result
\eq{
  \label{externcoh} 
  {\rm Ext}^0 (\i_* L_a, \i_* L_b)&= H^0(\Sigma, L_a\otimes L^\vee_b),  \\ 
  {\rm Ext}^1 (\i_* L_a, \i_* L_b)&= H^1(\Sigma, L_a\otimes L^\vee_b)+ 
  H^0(\Sigma, L_a\otimes L^\vee_b\otimes N_D),  \\  
  {\rm Ext}^2 (\i_* L_a, \i_* L_b)&= H^2(\Sigma, L_a\otimes L^\vee_b)+ 
  H^1(\Sigma, L_a\otimes L^\vee_b\otimes N_D),  \\  
  {\rm Ext}^3 (\i_* L_a, \i_* L_b)&= \phantom{aaaaaaaaaaaaaaa}\  
  H^2(\Sigma, L_a\otimes L^\vee_b\otimes N_D). 
}
For branes wrapping two different 4-cycles intersecting over a curve
$C=\Sigma_a\cap \Sigma_b$,
one has to compute the cohomology classes
\eq{
           H^i(C, L^\vee_a\otimes L_b\otimes K^{1\over 2}_C)\,, \qquad i=0,1\,.
}
For more details, we refer to \cite{Blumenhagen:2008zz}.

Moreover, the $U(1)$ factors in the total gauge group are often
anomalous. In string theory, these anomalies are cancelled by 
the Green-Schwarz mechanism, where a shift symmetry
of an axion gets gauged so that it becomes the longitudinal
mode of a massive abelian vector field. The abelian gauge symmetry
gets broken but survives as a perturbative global symmetry
that is only broken by D-brane instantons. We recall that 
an abelian gauge field can even gain a mass without having
an anomaly.

Let us consider now the tree-level gauge couplings at the
string scale 
\eq{
\label{gaugecoupling}
      \kappa_a  {4\pi\over g_a^2} = \tau_a -{1\over 2 g_s}\, \, \int_{\Sigma} c^2_1({\cal L}_a)
}
where $\tau_a ={1\over 2\, \ell_s^4}\int_{\Sigma_a}  J\wedge J $
denotes  the volume of the four-cycle (in Einstein frame)
in units of the string length $\ell_s$ and   $g_s=e^\varphi$ is 
the string coupling constant.
Here $\kappa_a=1$ for a $U(N)$ stack and $\kappa_a=2$ for an
$SP(2N)/SO(2N)$ stack\footnote{The factor $\kappa=1,2$ is due to the
``doubled'' embedding of $U(N)$ into $SO(2N)/SP(2N)$ 
$T_{2N}={1\over \sqrt{2}}\left(\begin{matrix} t_N & 0 \\ 0 & -t^T_N \end{matrix}\right)$.}.
As usual  in string theory, these couplings will receive
one-loop threshold corrections.
Note that for all branes wrapping the same 4-cycle, 
the differences among the gauge couplings $g_a$ come
only from the line bundles and in fact only from the value 
of $\int_{\Sigma_a} c^2_1({\cal L}_a)\in \mathbb Z$ \footnote{
These values also depend on the twist line bundle ${\cal R}_a$.}.

In contrast to single GUT groups, the gauge couplings at the string scale
are not necessarily equal. This situation and its consequences for
gauge coupling unification  for intersecting
D-brane models and F-theory models have been discussed in  \cite{Blumenhagen:2003jy}.
However, one can still design
situations where precise unification can occur.

If all branes are of $U(N)$ type, one can wrap all stacks around 
a single 4-cycle $\Sigma$ of size $\tau$ and distinguish them solely by
the line-bundles ${\cal L}_a$.
\begin{itemize}
\item{In the regime $\tau\gg g_s^{-1}>1$ with the differences of the 
  flux integrals not being too large, the gauge couplings approximately
   unify with
   ${4\pi\over g_a^2} = \tau$ where the flux dependent corrections can
   be considered as small threshold corrections. The values of 
   $\tau$ and $g_s$ are determined by moduli stabilization.}
\item{Since the second term in \eqref{gaugecoupling} only depends
    on the topological quantity ${\rm ch}_2({\cal L}_a)$, all gauge
    couplings can still be degenerate even if the ${\cal L}_a$
    themselves are different.}
\end{itemize}

\noindent
If some of the branes are of $SP$-type then due to the factor
$\kappa=2$ in \eqref{gaugecoupling}, the $SP$ branes
should wrap a  different four-cycle $\Sigma_{SP}$  than the $U(N)$
branes $\Sigma_{U}$ with $\tau_{SP}=2\tau_{U}$. In this case,
one also gets approximate gauge coupling unification in the sense
just explained  for $U(N)$ stacks.

Thus, in the following we will work in a scheme where we realize 
the LRSM on such intersecting D7-branes and we will assume that
we have gauge coupling unification at the string or unification scale
up to small threshold corrections. Moreover, the initial brane
realization should be supersymmetric at the string scale,
where, as usual, supersymmetry breaking will be mediated to the
observable sector by generating soft masses of the order $M_{\rm
  SUSY}$.

\section{Brane realizations of LRSM}

In this section two principal realizations of the LRSM in terms of
intersecting D7-branes are presented. We do not provide  fully
fledged global string compactifications, but instead only local brane
configurations that satisfy the consistency conditions following
from string theory. The realization of the SM itself in terms of intersecting branes
is also known as the Madrid quiver, first presented in \cite{Ibanez:2001nd}.
That model is very similar to  the LRSM quivers \cite{Blumenhagen:2000ea} to be discussed
below. 
%Similar LRSM brane configurations were considered very early in
%\cite{Blumenhagen:2000ea}.

\subsubsection*{LRSM quiver A}

First we consider the simplest quiver A, in which the
$SU(2)_L\times SU(2)_R$ sector is realized on orientifold
invariant branes supporting $SP(2)\simeq SU(2)$ gauge group.
We introduce four stacks of D7-branes carrying appropriate line
bundles such that one gets the initial gauge symmetry
$U(3)\times SP(2)\times SP(2) \times U(1)$ and that
the massless spectrum  is the one shown  in table \ref{tablequiverA}.
\begin{table}[ht]
  \centering
  \begin{tabular}{c|c||c|c|c|c||c}
    number      & field             &  $SU(3)$ & $SU(2)_L$   & $SU(2)_R$ &
    $U(1)_a\times U(1)_d$ & $U(1)_{B-L}$\\
    \hline \hline
             &   & &       &    &  & \\[-0.4cm]
      $3$ & $Q_L$    & $3$ & $2$ & $1$ & $(1,0)$  & $1/3$\\
      $3$  & $(Q_R)^c$   & $\ov 3$ & $1$ & $2$ & $(-1,0)$  & $-1/3$\\
      $3$   & $\ell_L$    & $1$ & $2$ & $1$ & $(0,1)$  & $-1$\\
     $3$     & $(\ell_R)^c$   & $1$ & $1$ & $2$ & $(0,-1)$  & $1$\\
    \hline 
    $N_\Phi$     & $\Phi$  & $1$ & $2$ & $2$ & $(0,0)$  & $0$\\
    $N_H$     & $H_R^u$  & $1$ & $1$ & $2$ & $(0,1)$  & $-1$\\
    $N_H$      & $H_R^d$  & $1$ & $1$ & $2$ & $(0,-1)$  & $1$\\
 \hline 
   $N_\Delta$      & $\Delta_0$  & $1$ & $1$ & $3$ & $(0,0)$  & $0$\\
      $N_S$     & $S^u$  & $1$ & $1$ & $1$ & $(0,2)$  & $-2$\\
    $N_S$      & $S^d$  & $1$ & $1$ & $1$ & $(0,-2)$  & $2$
  \end{tabular}
  \caption{\small Massless left-handed  spectrum of LRSM quiver A.}
  \label{tablequiverA}
\end{table}

One has two abelian gauge factors $U(1)_B={1\over 3} U(1)_a\subset U(3)$ and
$U(1)_L=U(1)_d$, whose charges can be identified with baryon and
lepton number, respectively. However, the only anomaly-free
combination is $U(1)_{B-L}={1\over 3}U(1)_a-U(1)_d$. Therefore, the
orthogonal combination receives a mass via the Green-Schwarz
mechanism. This in particular means that baryon and lepton number
survive as global symmetries and can protect the proton from decaying.
In particular, the unification scale can be smaller than the usual
GUT-scale  $M_{\rm  GUT}=2\cdot 10^{16}\,$GeV.

The gauge coupling of a linear
combination $U(1)_{B-L}=\sum c_i\, U(1)_i$  with  $U(1)_i\subset U(N_i)$
can be computed as
\eq{
\label{couplingcombi}
           {1\over \alpha_{B-L}}=\sum_i {1\over 2} N_i\, c_i^2\, {1\over
             \alpha_i}={1\over2}\left( {1\over 3} {1\over
                 \alpha_a}+{1\over \alpha_d}\right)={2\over 3} {1\over
               \alpha_s}\,,
}
%\eq{
%             {1\over \alpha_{B-L}}={1\over2}\left( {1\over 3} {1\over
%                 \alpha_a}+{1\over \alpha_d}\right)={2\over 3} {1\over
%               \alpha_s}
%}
assuming  stringy gauge coupling unification, i.e. $\alpha_s=\alpha_a=\alpha_d$. 
Note that this is the same relation as for $SO(10)$ GUTs.

The bi-doublet Higgs field $\Phi$ originates from a vector-like
intersection between the two $SP(2)$-branes. As indicated in the
table,  there could be more than just a single such  Higgs field, 
but its minimal non-vanishing number is really $N_\Phi=1$.
It is clear that for intersecting branes, one cannot get an $SU(2)_R$
triplet with  $Q_{B-L}=\pm 2$. The open string of such a
massless mode would need four instead of two ends.
Therefore, the breaking of the $SU(2)_R\times U(1)_{B-L}$ gauge
symmetry to $U(1)_Y$ has to proceed via a Higgs field  in the
doublet representation of $SU(2)_R$ with $Q_{B-L}=1$.
Note that anomaly cancellation  forces us here to
introduce such Higgs fields in vector-like pairs $H_R^{u,d}$.
We also added vector-like pairs $S^{u,d}$ 
of fields transforming in the symmetric representation of $U(1)_d$
and $SU(2)_R$ triplets with $Q_{B-L}=0$. 
Note that in contrast to other approaches,  parity symmetry 
$P:SU(2)_L\leftrightarrow SU(2)_R$ is 
broken explicitly in this Higgs sector.

%\vspace{1cm}
\subsubsection*{LRSM quiver B}

One can also realize  the $SU(2)_{L,R}$ gauge symmetries on $U(2)$
type of branes. 
In this case one has four $U(1)$ factors, of which we assume that 
only the anomaly-free combination $U(1)_{B-L}={1\over 3}U(1)_a-U(1)_d$ 
stays massless after the Green-Schwarz mechanism has been employed.
A configuration consistent with the stringy constraints is presented
in table \ref{tablequiverB}. Note that the generation of all possible
SM Yukawa couplings and anomaly cancellation forces one to introduce
an  even number of  bi-doublet 
Higgses $\Phi$.
\begin{table}[ht]
  \centering
  \begin{tabular}{c|c||c|c|c|c||c}
    number      & field             &  $SU(3)$ & $SU(2)_L$   & $SU(2)_R$ &
    $U(1)^4$ & $U(1)_{B-L}$\\
    \hline \hline
             &   & &       &    &  & \\[-0.4cm]
      $2$ & $Q_L$    & $3$ & $2$ & $1$ & $(1,1,0,0)$  & $1/3$\\
      $1$ & $Q_L$    & $3$ & $2$ & $1$ & $(1,-1,0,0)$  & $1/3$\\
      $2$  & $(Q_R)^c$   & $\ov 3$ & $1$ & $2$ & $(-1,0,1,0)$  &
      $-1/3$\\
      $1$  & $(Q_R)^c$   & $\ov 3$ & $1$ & $2$ & $(-1,0,-1,0)$  & $-1/3$\\
      $3$   & $\ell_L$    & $1$ & $2$ & $1$ & $(0,-1,0,1)$  & $-1$\\
     $3$     & $(\ell_R)^c$   & $1$ & $1$ & $2$ & $(0,0,-1,-1)$  & $1$\\
    \hline 
    $N_\Phi/2$     & $\Phi^u$  & $1$ & $2$ & $2$ & $(0,1,1,0)$  & $0$\\
    $N_\Phi/2$     & $\Phi^d$  & $1$ & $2$ & $2$ & $(0,-1,-1,0)$  & $0$\\
    $N_H$     & $H_R^u$  & $1$ & $1$ & $2$ & $(0,0,1,1)$  & $-1$\\
    $N_H$      & $H_R^d$  & $1$ & $1$ & $2$ & $(0,0,-1,-1)$  & $1$\\
\hline 
    $N_\Delta/2$     & $\Delta_0^u$  & $1$ & $1$ & $3$ & $(0,0,2,0)$  & $0$\\
    $N_\Delta/2$     & $\Delta_0^d$  & $1$ & $1$ & $3$ & $(0,0,-2,0)$  & $0$\\
  $N_S$     & $S^u$  & $1$ & $1$ & $1$ & $(0,0,0,2)$  & $-2$\\
    $N_S$      & $S^d$  & $1$ & $1$ & $1$ & $(0,0,0,-2)$  & $2$
  \end{tabular}
  \caption{\small Massless left-handed  spectrum of  LRSM quiver B.
    Anomaly  cancellation/tadpole cancellation enforces $N_{\Phi}$ even.}
  \label{tablequiverB}
\end{table}

In the next section, we analyze
the running of the four gauge couplings in the regime
$M_R<M_{\rm SUSY}<M_{U}$. Even though the matter content
of the models was motivated by D-brane constructions,
the upcoming analysis only depends on the former
and is therefore generically valid.

\section{Gauge Coupling Unification}

%For the models presented in the previous section we now analyze
%the running of the four gauge couplings in the regime
%$M_R<M_{\rm SUSY}<M_{U}$. 
We first run the 
Standard Model couplings from the weak scale up to the new left-right
unification scale $M_R\sim 2\,$TeV. 
For the values of the gauge couplings at the weak scale $M_Z=91.18\,$GeV we 
took  $\alpha_s=0.1172$, 
$\alpha=1/127.934$ and  $\sin \theta_w=0.23113$. Then, at the scale
$M_R$ one obtains
\eq{
   \alpha_s(M_R)=0.0835\,,\quad \alpha_L(M_R)=0.0321\,,\qquad
   \alpha_Y(M_R)=0.0105\,.
}
At the scale $M_R$ the hypercharge coupling
splits into the $SU(2)_R$ and the $U(1)_{B-L}$ coupling according
to 
\eq{    
\label{hyperbminusl}
                   {1\over \alpha_Y}={1\over \alpha_R}+{1\over
                     \alpha_{B-L}}\,.
}
The running beyond $M_R$ is evaluated  under the following
two assumptions
\begin{enumerate}
\item Following the extended survival hypothesis
  \cite{delAguila:1980at},  in the regime $M_R<\mu<M_{\rm SUSY}$  there is just the minimal
   particle content of the non-supersymmetric LRSM, i.e.
   in particular one scalar Higgs bi-doublet $\Phi$ and one scalar Higgs
    doublet $H_R^u$. Due to supersymmetry breaking, they are
    expected to gain soft masses of the order of $M_{\rm SUSY}$.
    Since for  $M_{\rm SUSY} \gg M_R$ one cannot refer to supersymmetry
    to solve the hierarchy  problem, one needs some fine-tuning or string landscape
    argument to achieve this.

\item In the regime $M_{\rm SUSY}<\mu<M_{U}$ all the 
  supersymmetric states that the intersecting D-brane model
  provides contribute to the running. This includes  $N_\Phi$ chiral fields
 in  the bi-doublet representation, as well as  
 $N_H$ vector-like  Higgs fields $H_R$. We also allow for  $N_\Delta$ vector-like  fields
 $\Delta_0$  and $N_S$ vector-like  fields $S$.
\end{enumerate}

\noindent
The one-loop running of the four  gauge couplings occurs according to
\eq{
\label{running}
    {1\over  \alpha_i(\mu)} =  {1\over \alpha_i(M_R)} + {b_i\over 2\pi} 
                  \log\left({M_{\rm SUSY}\over M_R}\right)+
          {\tilde b_i \over 2\pi} \log\left( 
        {\mu\over M_{\rm SUSY} } \right) 
}
where $(\alpha_3,\alpha_2^L,\alpha_2^R,\alpha_1)=(\alpha_s,\alpha_L,\alpha_R,{2\over
  3}\alpha_{B-L})$
and the $\beta$-function coefficients in the non-super\-sym\-metric and
supersymmetric region are given by\footnote{Due to the normalization
  in \eqref{couplingcombi}, one has e.g. $\tilde b_{1}={3\over
    2}\sum_{i,{\rm chiral\,}}  {Q_{B-L,i}^2\over 4} $.}
\eq{
\label{fluxzerocomp}
\arraycolsep2pt
\begin{array}{lcl@{\hspace{70pt}}lcl}
b_3&=&7\,, & \tilde b_3 &=& 3\,,\\[5pt]
b_2^L&=&3\,, & \tilde b_2^L &=& -N_\Phi\,,\\[5pt]
b_2^R&=&-{1\over 6} n_H^l+3\,, & \tilde b_2^R &=& -N_H-N_\Phi-2N_\Delta\,,\\[5pt]
b_{1}&=&-{1\over 4} n_H^l-4\,, & \tilde b_{1} &=& -{3\over 2}N_H-3
N_S-6\,.
\end{array}
}
Here we were leaving the number $n_H^l$ of light scalar Higgs fields
in the doublet representation of $SU(2)_R$   after supersymmetry
breaking as an open parameter. A fermion from the superfield $H_R$
that remains light contributes like $n_H^l=2$.
The extended survival hypothesis  means that we have $n_H^l=1$.

Taking the relation \eqref{hyperbminusl} into account, we have three unspecified
parameters namely $\alpha_R(M_R)$, $M_{\rm SUSY}$ and $M_{U}$
that can be uniquely determined by requiring that all four
gauge couplings unify at a  scale $M_{U}$.
For self-consistency one needs
%to eventually check whether the assumed hierarchy 
$M_R<M_{\rm SUSY}<M_{U}$  and that  all
couplings remain in the perturbative regime.
%As we will see this won't be the case for all choices of the number  $N_H$. 
Let us discuss the two classes $N_{\Phi}=$even/odd
%intersecting brane scenarios
separately.
% where for quiver A we choose the minimal value $N_\Phi=1$
%and for quiver B $N_\Phi=2$.

\subsubsection*{LRSM quiver$\,$A for odd $N_{\Phi}$}

For $N_\Delta=N_S=0$, choosing different values of the number of Higgs fields $N_{\Phi}$
and $N_H$  and 
solving for  $g_R$, $M_{\rm SUSY}$ and $M_{U}$, for quiver A
one obtains the scales shown in table \ref{tablescalesA}.
\begin{table}[ht]
  %\centering
  \begin{tabular}{c|c|c}
    $N_H$                   &  $M_{\rm SUSY}$ [GeV] & $g_R$    \\
    \hline \hline
                      &    &  \\[-0.4cm]
    $<3$       & $<M_R$ &   \\[0.1cm]
       $3$       & $19.2\cdot 10^{3}\,$ & 0.532   \\
      $4$       & $1.26\cdot 10^{8}\,$ & 0.537   \\
       $5$       & $1.21\cdot 10^{10}\,$ & 0.540   \\
       $6$       & $ 1.99\cdot 10^{11}\,$ & 0.542   \\
    $\downarrow$ &  & \\
    $\infty$ & $M_{\rm GUT}$  &  $\sim 0.55$ 
  \end{tabular} \hspace{1.2cm}
   \begin{tabular}{c|c|c|c}
    $N_H$                   &  $M_{\rm SUSY}$ [GeV] & $g_R$  & $M_{U}$ [GeV] \\
    \hline \hline
                      &    &  \\[-0.4cm]
    $<6$       & $<M_R$ &   \\[0.1cm]
       $6$       & $19.2\cdot 10^{3}\,$ & 0.507  & $2.18\cdot 10^{12}$ \\
      $7$       & $1.34\cdot 10^{7}\,$ & 0.516  & $1.94\cdot 10^{13}$  \\
       $10$       & $7.70\cdot 10^{8}\,$ & 0.522 & $7.47\cdot 10^{13}$   \\
       $12$       & $ 1.21\cdot 10^{10}\,$ & 0.526 & $1.87\cdot 10^{14}$   \\
    $\downarrow$ &  & \\
    $\infty$ & $M_{\rm GUT}$  &  $\sim 0.55$ & $M_{\rm GUT}$
  \end{tabular}
  \caption{\small $M_{\rm SUSY}$, $g_R$ and $M_U$  for different
   values of $N_H$ with $N_\Delta=N_S=0$. 
   The left table is for $N_{\Phi}=1$ ($M_U=M_{\rm GUT}$) and the right one for $N_{\Phi}=3$. }
  \label{tablescalesA}
\end{table}

\noindent 
For $N_H\le 3,6$ one gets $M_{\rm SUSY}<M_R$ which is in conflict
with the assumption. The lowest possible values are therefore $N_H=3,6$, for
which the supersymmetry breaking scale comes out one order of
magnitude larger than $M_R$ and is actually the same for
the two choices $N_{\Phi}=1$ and $N_{\Phi}=3$. 

\begin{figure}[ht]
\begin{center} 
\includegraphics[width=0.7\textwidth]{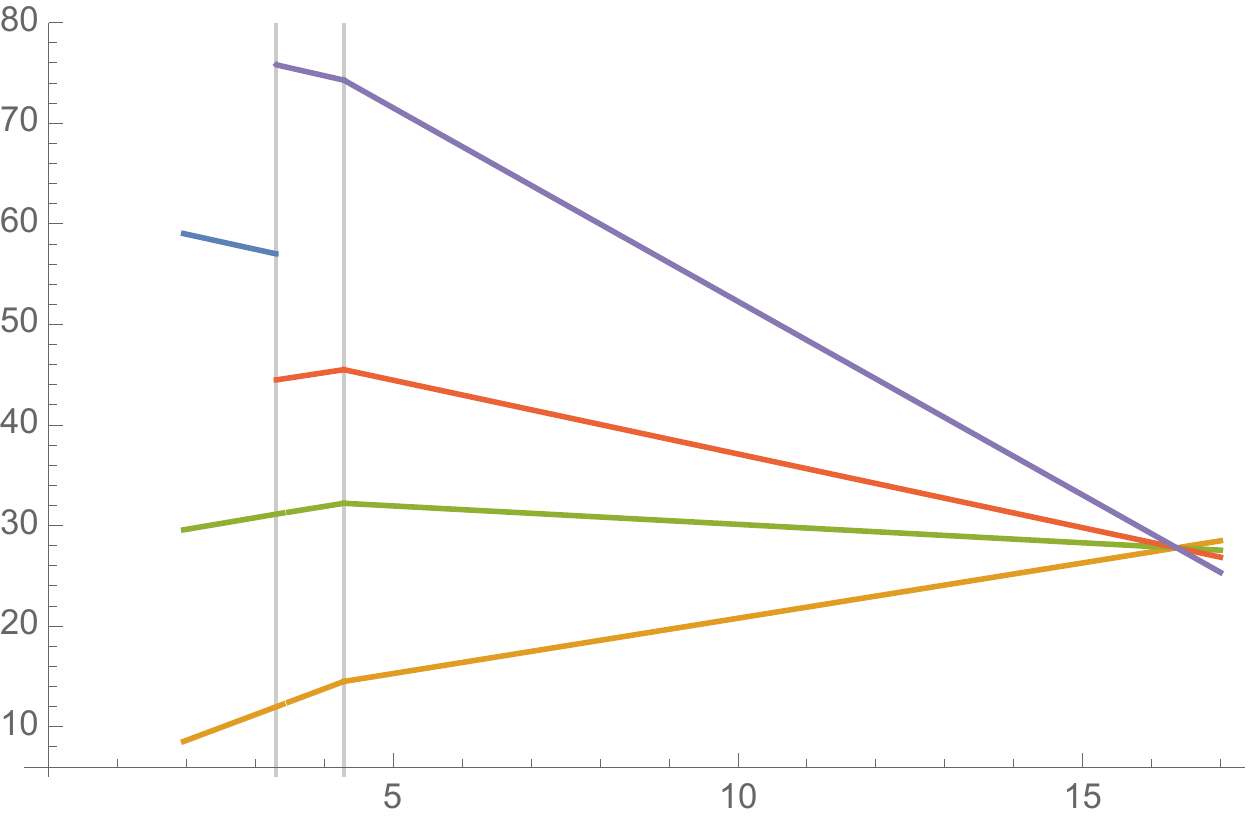}
\begin{picture}(0,0)
  \put(0,9){${\rm log}_{10}(\mu)$}
   \put(-276,133){${3\over 5}\alpha_Y^{-1}$}
   \put(-270,65){$\alpha_L^{-1}$}
   \put(-270,18){$\alpha_s^{-1}$}
   \put(-250,100){$\alpha_R^{-1}$}
   \put(-264,173){${3\over 2}\alpha_{B-L}^{-1}$}
%   \put(-145,125){$M_{3\overline 3}$}
%   \psline[linewidth=1pt, linestyle=dotted]{-}(-4.75,0.3)(-4.75,4.2)

\end{picture}
\hspace*{15pt}
\end{center} 
\vspace{-15pt}
\caption{{\small 
One-loop running of the LRSM gauge couplings for $N_{\Phi}=1$, $N_H=3$, $N_\Delta=N_S=0$
and $n_H^l=1$  with $M_{\rm SUSY}=19.2\,$TeV and $g_R=0.532$.} 
\label{figrunningA}}
\vspace*{-5pt}
\end{figure}
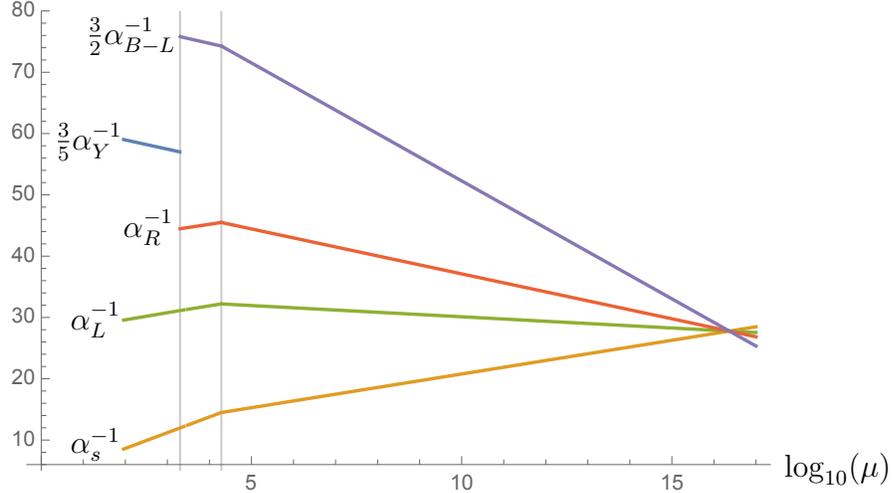

This value would be out of reach of the LHC
Run-2,
but threshold corrections at the high scale and two-loop corrections
to the running might lower this value. This issue will be  discussed
below. The value of the $SU(2)_R$ gauge coupling 
does only  vary slightly   in the region $0.5<g_R<0.55$.
Recall that baryon number is still a perturbative global
symmetry so that for $M_U<M_{\rm GUT}$ fast proton decay is not an issue.
In figure \ref{figrunningA}, the running coupling constants are shown
for $N_{\Phi}=1$ and $N_H=3$.
\vspace{0.3cm}
%%%%%%%%%%%%%%%%%%%%%%%%%%%%%%%%%%%%%%%%%%%%%%%
%%%%%%%%%%%%%%%%%%%%%%%%%%%%%%%%%%%%%%%%%%%%%%%
%%%%%%%%%%%%%%%%%%%%%%%%%%%%%%%%%%%%%%%%%%%%%%%
%%%%%%%%%%%%%%%%%%%%%%%%%%%%%%%%%%%%%%%%%%%%%%%
%%%%%%%%%%%%%%%%%%%%%%%%%%%%%%%%%%%%%%%%%%%%%%%

We observed that the minimal value of 
the supersymmetry breaking scale $M_{\rm SUSY}=19.2\,$TeV
was appearing for both choices  of the number of  bi-doublet Higgses
$N_\Phi=1,3$. Clearly, this asks for an explanation.
First, one notices that the values of $M_{\rm SUSY}$  and $M_U$ do 
only depend on the combination $(N_H+N_\Delta+N_S)$ with only $g_R$ depending
on their individual values.
For the (minimal) 
choice $(N_H+N_\Delta+N_S)^{\rm min}={3\over 2}(N_\Phi+1)$,
the supersymmetry breaking scale can be generically determined as
\eq{
      M_{\rm SUSY}=M_R\, \exp\left[ {2\pi\over 14-n_H^l}\left({12\over
            \alpha_L(M_R)}-{7\over \alpha_s(M_R)}-{3\over
            \alpha_Y(M_R)}\right)\right]\,.
}
Surprisingly, the scale does not depend on $N_\Phi$, but
only on the LR-breaking scale $M_R$  and the number of light Higgses.
For $n_H^l=1$ one gets  $M_{\rm SUSY}=19.2\,$TeV, which increases
for larger values of $n_H^l$. For $M_R=1.8\,$TeV one finds $M_{\rm SUSY}=16\,$TeV.

One can show that for $N_{\Phi}\le 21$, the number $(N_H+N_\Delta +N_S)^{\rm min}={3\over 2}(N_\Phi+1)$ is
indeed the minimal threshold value  guaranteeing  $M_{\rm SUSY}>M_R$.
 It is in this sense that, for $M_R=2\,$TeV,  this is a universal
 result. For $N_{\Phi}=21$, $N_\Delta =N_S=0$   and $n_H^l=1$ one obtains
 $M_{\rm SUSY}=19.2\, {\rm TeV}$, $g_R=0.476$ and $M_U=1.98\cdot 10^{6}\,{\rm GeV}$,
hence featuring  smaller values of the $SU(2)_R$ gauge coupling and the 
unification scale.

Taking the minimal choice $N_\Phi=1$, for all
possible partitions $(N_H+N_\Delta +N_S)^{\rm min}=3$ with $N_H\ge 1$
we find $M_{\rm SUSY}=19.2\,$TeV, $M_U=M_{\rm GUT}=2.3\cdot
10^{16}\,$GeV and the values of $0.48<g_R<0.6$ shown
in table \ref{tablesgR}. 

\begin{table}[ht]
  \centering
  \begin{tabular}{c|c}
    $(N_H,N_\Delta,N_S)$ & $g_R$   \\
    \hline \hline
                      &  \\[-0.4cm]
     $(3,0,0)$ &     $0.532$      \\
     $(2,1,0)$ &     $0.507$      \\
    $(2,0,1)$ &     $0.560$      \\
    $(1,2,0)$ &     $0.485$      \\
   $(1,1,1)$ &     $0.532$      \\
   $(1,0,2)$ &     $0.594$      
  \end{tabular}
  \caption{\small Values of  $g_R$ for  $N_\Phi=1$ and $(N_H+N_\Delta +N_S)^{\rm min}=3$.}
  \label{tablesgR}
\end{table}

%%%%%%%%%%%%%%%%%%%%%%%%%%%%%%%%%%%%%%%%%%%%%%%

\subsubsection*{LRSM quivers$\,$A,B for even $N_{\Phi}$}

The same computation can be performed for an even number of 
bi-doublet Higgs fields. Recall that for quiver B, anomaly
cancellation enforced these fields to come in  vector-like pairs.
In table \ref{tablescalesB} we list the  resulting mass scales
for various choices of $N_\Phi$ for the corresponding minimal value
of $N_H$ and $N_\Delta=N_S=0$.
\begin{table}[ht]
  \centering
  \begin{tabular}{c|c|c|c|c}
    $N_\Phi$ & $N^{\rm min}_H$              &  $M_{\rm SUSY}$ [GeV] &
    $g_R$   & $M_{U}$ [GeV] \\
    \hline \hline
                      & &    &  & \\[-0.4cm]
     $2$ &     $5$       & $1.48\cdot 10^{6}$  & 0.521  &  $ 2.12\cdot
       10^{14}\,$GeV \\
    $4$ &     $8$       & $4.91\cdot 10^{5}$  & 0.506  &  $ 6.19\cdot
       10^{11}\,$GeV \\
   $6$ &     $11$       & $2.56\cdot 10^{5}$  & 0.497  &  $ 1.90\cdot
       10^{10}\,$GeV \\
$8$ &     $13$       & $1.5\cdot 10^{3}$  & 0.483  &  $ 9.40\cdot
       10^{7}\,$GeV 
  \end{tabular}
  \caption{\small Values of the $M_{\rm SUSY}$, $g_R$ and $M_{U}$ for different
   values of $N_{\Phi}$ and the corresponding minimal  values of
   $N_H$ with $n_H^l=1$ and $N_\Delta=N_S=0$.}
  \label{tablescalesB}
\end{table}

\noindent
Here, we do not find the same universality as for $N_\Phi$ odd.
This would arise  for half-integer values of the number of vector-like
Higgs fields $H_R$. 
%For the first even values of $N_\Phi$ the 
%various mass scales are displayed for the minimal choices
%of $N_H$. 
For $2\le N_\Phi\le 6$,  the minimal choices
of $N_H$ are parameterized as
$N^{\rm min}_H={3\over 2}N_\Phi+2$. For $N_\Phi=8$ the branch changes
to  $N_H^{\rm min}={3\over 2}N_\Phi+1$. 

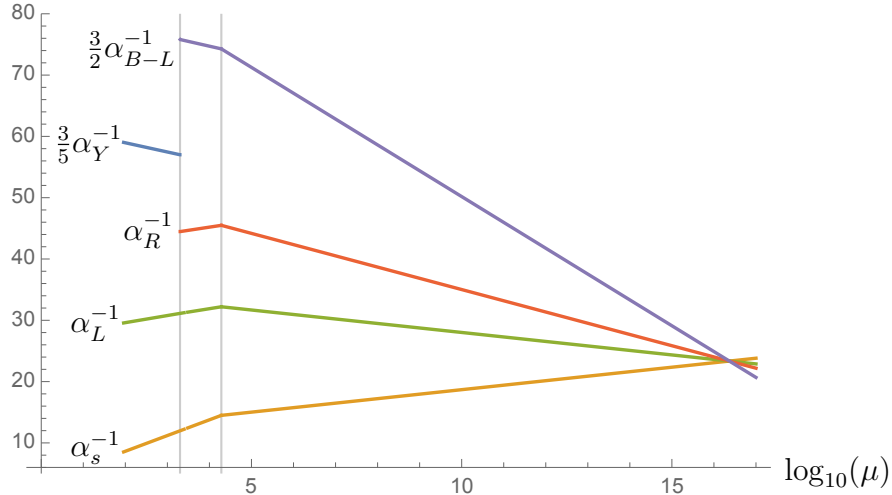
\begin{figure}[ht]
\begin{center} 
\includegraphics[width=0.7\textwidth]{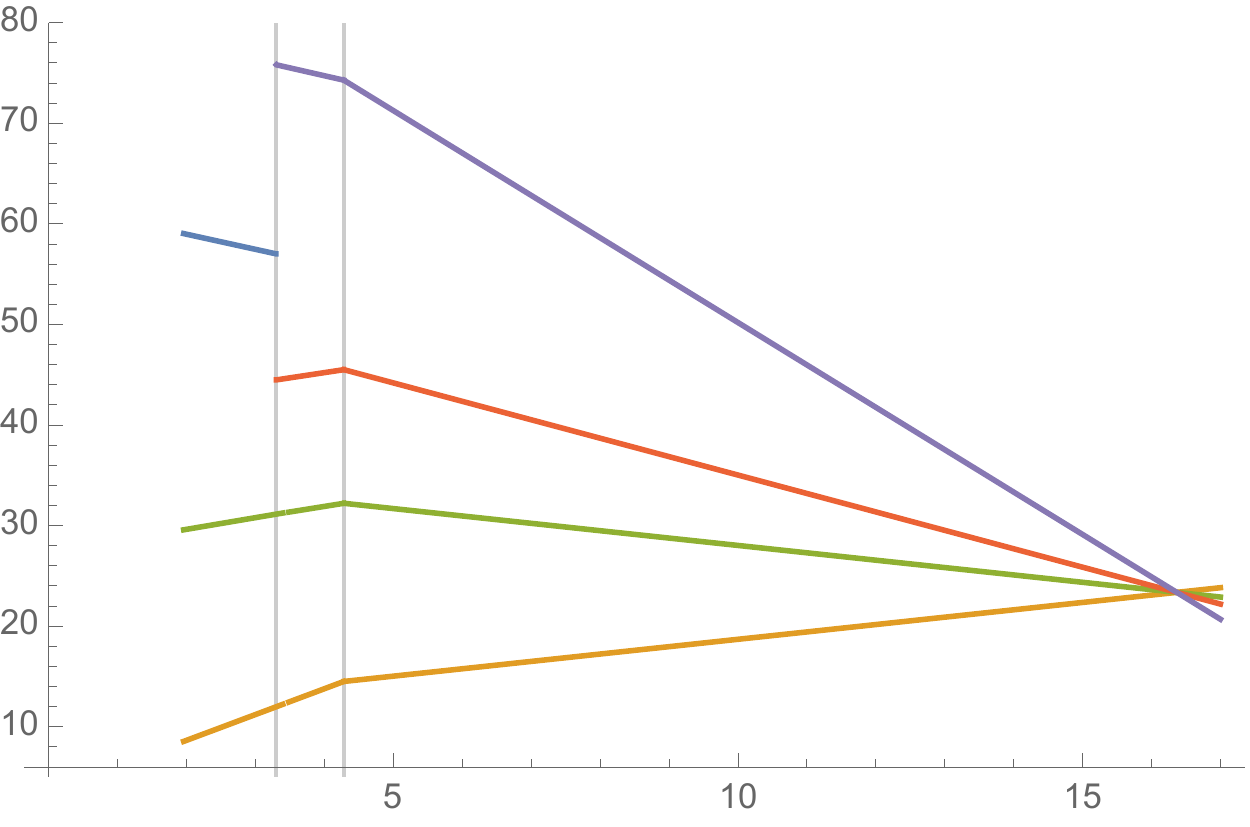}
\begin{picture}(0,0)
  \put(0,9){${\rm log}_{10}(\mu)$}
   \put(-276,133){${3\over 5}\alpha_Y^{-1}$}
   \put(-270,65){$\alpha_L^{-1}$}
   \put(-270,18){$\alpha_s^{-1}$}
   \put(-250,100){$\alpha_R^{-1}$}
   \put(-264,169){${3\over 2}\alpha_{B-L}^{-1}$}
%   \put(-145,125){$M_{3\overline 3}$}
%   \psline[linewidth=1pt, linestyle=dotted]{-}(-4.75,0.3)(-4.75,4.2)
\end{picture}
\hspace*{15pt}
\end{center} 
\vspace{-15pt}
\caption{{\small 
Running for $N_\Phi=2$, $N_T=1$, $N_\Delta=N_S=0$ and $N_H=3$
with $M_{\rm SUSY}=19.2\,$TeV and $g_R=0.532$. } \label{figrunningB}}
\vspace*{-5pt}
\end{figure}

What one can do though is to add $N_T$ vector-like pairs
of massless modes in the representation $(3,1,1,\pm 2/3)$.
These arise from open strings stretched between the 
$U(3)$ stack and the $U(1)_d$ stack.
The $\beta$-function  coefficients in the susy regime $M_R<\mu<M_U$  then read
\eq{
\label{fluxzerocompb}
\tilde b_3 &= 3-N_T\,,\\[1pt]
\tilde b_2^L &= -N_\Phi\,,\\[1pt]
\tilde b_2^R &= -N_H-N_\Phi-2 N_\Delta\,,\\[1pt]
\tilde b_{1} &= -{3\over 2}N_H-3
N_S-6-N_T\,.
}
Then it is clear that choosing $N_\Phi$ even and $N_T=1$, the three values 
$M_{\rm SUSY}$, $g_R(M_R)$ and $M_U$ remain the same as for 
$N_\Phi-1$ bi-doublet Higgses and $N_T=0$. Only the value of the unified
gauge coupling changes.
Hence, one is back to the discussion
for odd $N_\Phi$.
In figure \ref{figrunningB}, the one-loop running coupling constants are shown
for $N_\Phi=2$, $N_T=1$ and $N_H=3$. 
%\vspace{0.2cm}
%%%%%%%%%%%%%%%%%%%%%%%%%%%%%%%%%%%%%%%%%%%%%%%
%%%%%%%%%%%%%%%%%%%%%%%%%%%%%%%%%%%%%%%%%%%%%%%
%%%%%%%%%%%%%%%%%%%%%%%%%%%%%%%%%%%%%%%%%%%%%%%
%%%%%%%%%%%%%%%%%%%%%%%%%%%%%%%%%%%%%%%%%%%%%%%
%%%%%%%%%%%%%%%%%%%%%%%%%%%%%%%%%%%%%%%%%%%%%%%

 \subsubsection*{Comment}

The value of $M_{\rm SUSY}=19\,$TeV is a bit too large to be detected
directly at the LHC. However, our analysis was very strict in the sense
that we were assuming that all masses at  the scales $M_R$ and $M_{|rm SUSY}$
are the same, respectively.
Moreover, the computation
is performed only at  one-loop level, where two-loop corrections are usually
expected to give a 4\% correction to the couplings at the unification
scale. 
A correction  of the same order is expected if the gauge fluxes
on the stacks of branes lead to string threshold corrections, as
discussed in section 2.

\begin{figure}[ht]
\begin{center}
 \includegraphics[width=0.45\textwidth]{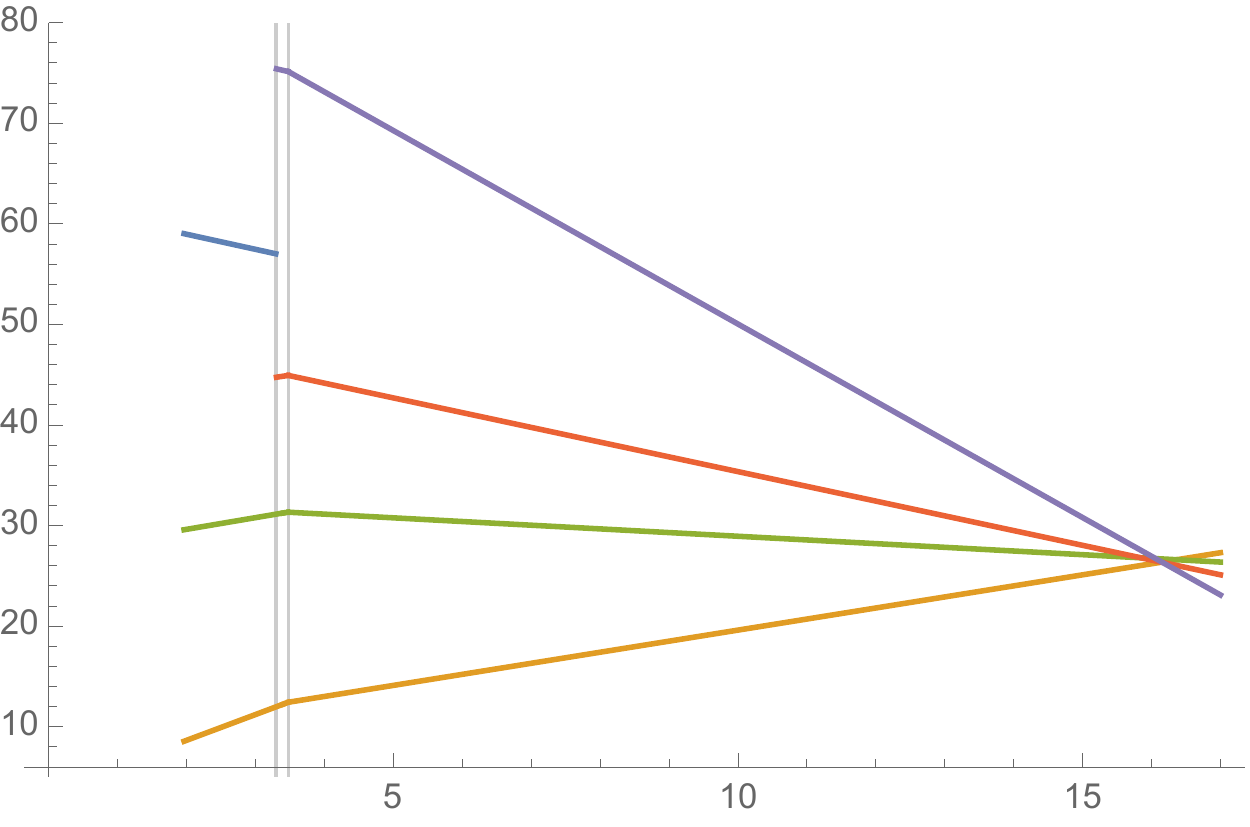}
\hspace{0.6cm}
\includegraphics[width=0.45\textwidth]{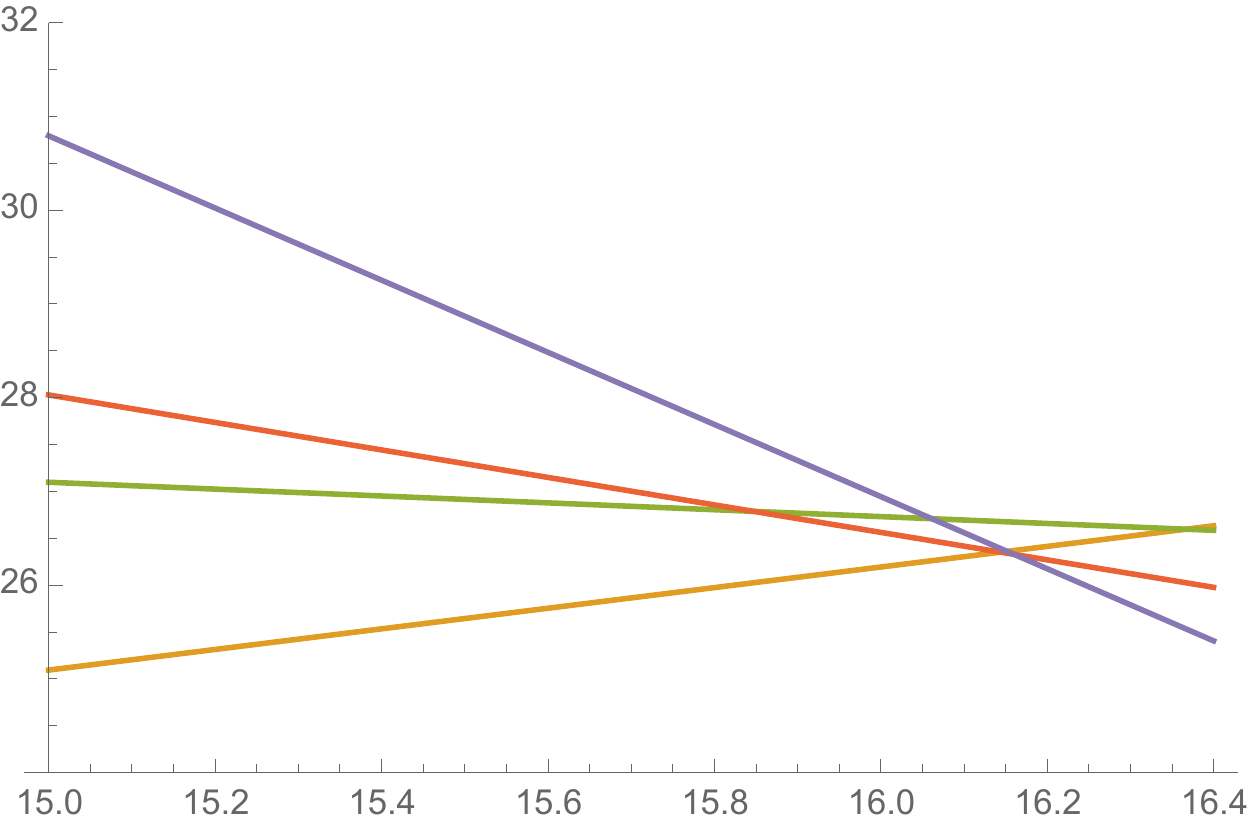}
%\begin{picture}(0,0)
%  \put(0,9){${\rm log}_{10}(\mu)$}
%   \put(-276,133){${3\over 5}\alpha_Y^{-1}$}
%   \put(-270,65){$\alpha_L^{-1}$}
%   \put(-270,18){$\alpha_s^{-1}$}
%   \put(-250,100){$\alpha_R^{-1}$}
%   \put(-264,169){${3\over 2}\alpha_{B-L}^{-1}$}
%   \put(-145,125){$M_{3\overline 3}$}
%   \psline[linewidth=1pt, linestyle=dotted]{-}(-4.75,0.3)(-4.75,4.2)
%\end{picture}
\hspace*{15pt}
\end{center} 
\vspace{-15pt}
\caption{{\small 
One-loop running of the LRSM gauge couplings for $N_H=3$  with
$M_{\rm SUSY}=3\,$TeV and $g_R=0.53$. } \label{figrunningC}}
\vspace*{-5pt}
\end{figure}

Just to get a first impression, for quiver A let us assume that e.g.
the supersymmetry breaking scale is $3\,$TeV, with $g_R=0.53$ and
just run the couplings up to the GUT scale. The result is shown in
figure  \ref{figrunningC}.

\vspace{0.2cm}
%%%%%%%%%%%%%%%%%%%%%%%%%%%%%%%%%%%%%%%%%%%%%%%
%%%%%%%%%%%%%%%%%%%%%%%%%%%%%%%%%%%%%%%%%%%%%%%
%%%%%%%%%%%%%%%%%%%%%%%%%%%%%%%%%%%%%%%%%%%%%%%
%%%%%%%%%%%%%%%%%%%%%%%%%%%%%%%%%%%%%%%%%%%%%%%
%%%%%%%%%%%%%%%%%%%%%%%%%%%%%%%%%%%%%%%%%%%%%%%
%%%%%%%%%%%%%%%%%%%%%%%%%%%%%%%%%%%%%%%%%%%%%%%

In the left picture one does not even see that the couplings do
not unify exactly. From the right picture one can estimate
a 2\% failure in doing so. Therefore, we conclude that for the minimal
value $N_H=3$, the inclusion of threshold corrections  and probably
also two-loop corrections can make a supersymmetry breaking scale
of just a bit above $M_R$ still consistent with stringy unification
at  around the GUT scale.

\section{Conclusions}

We presented possible realizations of LRSMs in terms
of intersecting D7-brane quivers. Since only 
 bifundamental and (anti-)symmetric 
representations can occur, the SM and LRSM
Higgs representations were fairly constrained. Employing this
point, we were considering the minimal matter content and
were studying the running coupling constants in the regime
between the LR-breaking scale of $2\,$TeV, as suggested 
by excesses in the LHC data, and a potential
unification scale.

We found that for a not too large number of bi-doublet Higgses
$H_\Phi$ and in each case a sufficiently large number of vector-like
fields  $\{H_R,\Delta_0,S\}$  one indeed achieves a  precise 
one-loop unification scenario in which 
the minimal value of the supersymmetry breaking scale
was determined as $19\,$TeV. This value 
was shown to be universal for an odd number of bi-doublet Higgs
fields. For an even number of such Higgses, by adding a 
further vector-like colour triplet state, the same universality could
be achieved\footnote{We were informed  that the model with 
$N_\Phi=2$, $N_T=1$, $N_H=1$ and $N_S=2$ has also been considered in \cite{Hirsch:2015fvq}.}.
Threshold corrections at the high scale might  lower this value so
that it can be detectable at the LHC Run-2.

Even though  one can contemplate various
variations and extensions of such stringy LRSM models, it is  satisfying
that with some stringy input and a number of reasonable  assumptions
it was possible to derive such a universal and in this scheme 
predictive result. Moreover, the value
of the gauge coupling constant $g_R$ was not varying much, either,  and
came out as $0.48<g_R< 0.6$. This is in the regime that was also 
suggested by a more
phenomenological analysis of the LRSM to fit the 
various, of course still not significant,  $2\,$TeV excesses observed
in Run-1 at the LHC.

It would be interesting to generalize the computation in various
directions like e.g.  to D-brane realizations of leptophobic models.
One should also  include  2-loop effects and consider threshold
effects both at the  small and the large scale.

\vspace{1cm}

\noindent
\emph{Acknowledgements:}
I thank Vic Feng,  Daniela Herschmann,  Dieter L\"ust and Florian Wolf  
for discussions and comments about the manuscript.
I also express my gratitude to the Erwin Schr\"odinger Institut (ESI) in Vienna
for hospitality.

\vspace{1cm}

%%%%%%%%%%%%%%%%%%%%%%%%%%%%%%%%%%%%%%%%%%%%%%%
%%%%%%%%%%%%%%%%%%%%%%%%%%%%%%%%%%%%%%%%%%%%%%%
%%%%%%%%%%%%%%%%%%%%%%%%%%%%%%%%%%%%%%%%%%%%%%%
%%%%%%%%%%%%%%%%%%%%%%%%%%%%%%%%%%%%%%%%%%%%%%%

\end{document}